\documentstyle[preprint,revtex,eqsecnum]{aps}

\begin{document}

\draft

\preprint{IFUSP/P-1072}
\preprint{September, 1993}

\begin{title}
Einstein-Cartan theory of gravity revisited
\end{title}

\author{Alberto Saa}
\begin{instit}
Instituto de F\'{\i}sica \\
Universidade de S\~ao Paulo, Caixa Postal 20516 \\
01498-970 S\~ao Paulo, SP \\
Brazil
\end{instit}

\begin{abstract}
The role of space-time torsion in general relativity is reviewed in
accordance with some
recent results on the subject. It is shown that, according to
the connection compatibility condition, the usual
Riemannian volume element is not appropriate in the presence of torsion. A
new volume element is proposed and used in the Lagrangian formulation for
Einstein-Cartan theory of gravity.
 The dynamical equations for the space-time geometry
and for matter fields are obtained, and some of
their new predictions and features are
discussed. In particular, one has that torsion propagates and that gauge fields
can interact with torsion without the breaking of gauge invariance. It is
shown also that the new Einstein-Hilbert action for Einstein-Cartan theory may
provide a physical interpretation for dilaton gravity in terms of the
non-riemannian structure of space-time.
\end{abstract}


\section{Introduction}

The Einstein-Cartan (EC) theory of gravity is a simple and natural
generalization of general relativity.
It was proposed in the early
twenties by \'Elie Cartan, who, before the introduction of the modern concept
of spin, suggested the possibility of relating space-time torsion to an
intrinsic angular momentum of matter. Nowadays, it is known that
 EC theory of gravity arises naturally
from the local gauge theory for the Poincar\'e's group, and that it is in
accordance with the available experimental data\cite{hehl,venzo}.
In such theory, space-time
is assumed to be a Riemann-Cartan (RC) manifold, and their dynamical
equations are gotten via a minimal action principle of the Einstein-Hilbert
type,
\begin{equation}
S_{\rm grav} = - \int d{\rm vol}\, R = - \int d^4x\sqrt{-g}\,R.
\end{equation}

Although in the macroscopic world space-time is usually assumed to be
torsionless, there are good reasons to believe that in the microscopic
level space-time must have a non-vanishing torsion\cite{hehl1}, and so,
microscopic
gravitational interactions should be described by EC theory. The
interest in such theories of gravity has grown in recent years also
due to their role in the
semi-classical description of quantum
fields on curved spaces, see for example \cite{odintsov} and references
therein.

In this work it will be shown that the usual Riemannian volume element,
which is used in the construction of the Einstein-Hilbert action, does
not play in RC space-times the same role that it does in a Riemannian
space-time. Endowed with an appropriate
volume element, EC theory will predict new effects.
The main new predictions are that torsion will propagate and that gauge
fields can interact with torsion without breaking of gauge invariance. It
is shown also that the proposed model can give a satisfactory physical
interpretation for the dilaton gravity that comes from string theory,
in terms of the non-riemannian structure of space-time.

The work is organized in four sections, where the first is this introduction.
In the Sect. 2, RC geometry is briefly introduced and the problem of
compatibility of volume elements is discussed. In the Sect. 3, the results
of Sect. 2 are used to propose some modifications to EC theory of gravity,
and the new equations for the vacuum is presented. Matter fields are also
considered in this section,
 namely, scalar, gauge, and fermion fields. In the last section,
it is shown that the proposed model requires some modifications in the
traditional point of view that only fermion fields can be source to the
non-riemannian structure of space-time, and further developments are discussed.

\section{RC manifolds and compatible volume elements}

The RC space-time ($U_4$) is a differentiable four dimensional
manifold endowed with a metric tensor $g_{\alpha\beta}(x)$ and with a
metric-compatible connection $\Gamma_{\alpha\beta}^\mu$, which is
non-symmetrical in its lower indices.
The following conventions will be adopted in this work:
${\rm sign}(g_{\mu\nu})=(+,-,-...)$,
$R_{\alpha\nu\mu}^{\ \ \ \beta} = \partial_\alpha\Gamma_{\nu\mu}^\beta
+ \Gamma_{\alpha\rho}^\beta\Gamma_{\nu\mu}^\rho -
(\alpha\leftrightarrow\nu)$, and
$R_{\nu\mu}=R_{\alpha\nu\mu}^{\ \ \ \alpha}$.
The anti-symmetric part of the
connection defines  a new tensor, the torsion tensor,
\begin{equation}
S_{\alpha\beta}^{\ \ \gamma} = \frac{1}{2}
\left(\Gamma_{\alpha\beta}^\gamma-\Gamma_{\beta\alpha}^\gamma \right).
\label{torsion}
\end{equation}
The metric-compatible connection can be written as
\begin{equation}
\Gamma_{\alpha\beta}^\gamma = \left\{_{\alpha\beta}^\gamma \right\}
- K_{\alpha\beta}^{\ \ \gamma},
\label{connection}
\end{equation}
where $\left\{_{\alpha\beta}^\gamma \right\}$ are
 the usual Christoffel symbols
from Riemannian space-time ($V_4$), and $K_{\alpha\beta}^{\ \ \gamma}$ is the
contorsion tensor, which is given in terms of the torsion tensor by
\begin{equation}
K_{\alpha\beta}^{\ \ \gamma} = - S_{\alpha\beta}^{\ \ \gamma}
+ S_{\beta\ \alpha}^{\ \gamma\ } - S_{\ \alpha\beta}^{\gamma\ \ }.
\label{contorsion}
\end{equation}
The connection (\ref{connection}) is used to define the covariant derivative of
a contravariant vector,
\begin{equation}
 D_\nu A^\mu = \partial_\nu A^\mu + \Gamma_{\nu\rho}^\mu A^\rho,
\label{covariant}
\end{equation}
and
it is also important to our purposes to introduce the covariant derivative of
a density $f(x)$,
\begin{equation}
D_\mu f(x) = \partial_\mu f(x) - \Gamma^\rho_{\rho\mu}f(x).
\end{equation}

The contorsion tensor (\ref{contorsion}) can be covariantly
split in a traceless part and in a trace,
\begin{equation}
K_{\alpha\beta\gamma} = \tilde{K}_{\alpha\beta\gamma} -
\frac{2}{3}\left(
g_{\alpha\gamma} S_\beta  - g_{\alpha\beta} S_\gamma
\right),
\label{decomposit}
\end{equation}
where $\tilde{K}_{\alpha\beta\gamma}$  is the traceless part and $S_\beta$ is
the trace of the torsion tensor, $S_\beta = S^{\ \ \alpha}_{\alpha\beta}$.
The $U_4$  curvature tensor is defined by using the
full connection (\ref{connection}), and it is given by:
\begin{equation}
R_{\alpha\nu\mu}^{\ \ \ \ \beta} = \partial_\alpha \Gamma_{\nu\mu}^\beta
- \partial_\nu \Gamma_{\alpha\mu}^\beta
+ \Gamma_{\alpha\rho}^\beta \Gamma_{\nu\mu}^\rho
- \Gamma_{\nu\rho}^\beta \Gamma_{\alpha\mu}^\rho .
\end{equation}
After some algebraic manipulations we get the following expression for the
scalar of curvature
\begin{equation}
R = g^{\mu\nu} R_{\alpha\mu\nu}^{\ \ \ \ \alpha} =
R^{V_4} - 4D_\mu S^\mu + \frac{16}{3}S_\mu S^\mu -
\tilde{K}_{\nu\rho\alpha} \tilde{K}^{\alpha\nu\rho},
\label{scurv}
\end{equation}
where $R^{V_4}$ is the $V_4$ scalar of curvature, calculated from the
Christoffel symbols.

In order to define a general covariant volume element in a manifold, it is
necessary to introduce a density quantity $f(x)$,
\begin{equation}
d^4x \rightarrow f(x) d^4x = d{\rm vol}.
\end{equation}
This is done
 in order to compensate
the Jacobian that arises from the transformation law
of the usual volume element
$d^4x$ under a coordinate transformation, and
usually, the density $f(x) = \sqrt{-g}$ is took to this purpose.

There are
natural properties that a volume element shall exhibit.
In $V_4$, the usual covariant volume element
\begin{equation}
d{\rm vol} = \sqrt{-g}\, d^4x,
\label{vele}
\end{equation}
is ``parallel'', in the sense that the scalar density $\sqrt{-g}$ obeys
\begin{equation}
D_\mu^{V_4}\sqrt{-g} = 0,
\end{equation}
where $D_\mu^{V_4}$ is the $V_4$ covariant derivative, defined using the
Christoffel symbols $\left\{_{\alpha\beta}^\gamma \right\}$. In mathematical
precise language, the volume element (\ref{vele}) is said to be compatible
with the connection in  $V_4$ manifolds.
One can infer that the volume element (\ref{vele}) is not ``parallel''
in $U_4$\cite{saa1}, since
\begin{equation}
D_\mu\sqrt{-g}= \partial_\mu\sqrt{-g} - \Gamma^\rho_{\rho\mu}\sqrt{-g} =
-2 S_\mu\sqrt{-g},
\end{equation}
as it can be checked using Christoffel symbols properties. This is the
main point that we wish to stress, it is the basic argument to our claim
that the usual volume element (\ref{vele}) is not appropriate in the
presence of torsion.

The question
that arises now is if it is possible to define a
 $V_4$-like, {\em i.e.} ``parallel'', volume element in $U_4$ manifolds.
In order to do it,
one needs to find out a  density $f(x)$ such that
$D_\mu f(x)=0$. Such density exists only if the trace
of the torsion tensor, $S_\mu$, can be obtained
from a scalar potential
\begin{equation}
S_\beta(x) = \partial_\beta \Theta(x),
\label{pot}
\end{equation}
and in this case we have $f(x)=e^{2\Theta}\sqrt{-g}$, and
\begin{equation}
d{\rm vol} = e^{2\Theta}\sqrt{-g} \,d^4x,
\label{u4volume}
\end{equation}
that is the ``parallel'' covariant $U_4$ volume element,
or in another words, the volume element (\ref{u4volume}) is compatible
with the connection in RC manifolds  obeying (\ref{pot}).

With
the volume element (\ref{u4volume}), we have the following generalized
Gauss' formula
\begin{equation}
\int d{\rm vol}\, D_\mu V^\mu =
\int d^4x \partial_\mu e^{2\Theta}\sqrt{-g} V^\mu =\
{\rm surface\ term},
\label{gauss}
\end{equation}
where we used
that
\begin{equation}
\label{gammacontr}
\Gamma^\rho_{\rho\mu}=\partial_\mu\ln e^{2\Theta}\sqrt{-g}
\end{equation}
under the hypothesis (\ref{pot}). It is easy to see that one cannot have a
generalized Gauss' formula of the type (\ref{gauss}) if the torsion does not
obey (\ref{pot}). We will return to discuss the role of the condition
(\ref{pot}) in the last section.

\section{Einstein-Cartan theory of gravity}

In this section, Einstein-Cartan theory of gravity will be reconstructed by
using the results of the Sect. 2. Space-time will be
 assumed to be a Riemann-Cartan
manifold with the ``parallel'' volume element (\ref{u4volume}), and of course,
it is implicit the restriction that the trace of the torsion tensor is
derived from a scalar potential, condition (\ref{pot}).
With this hypothesis, EC theory of gravity will predict new effects, and they
will be pointed out in the following subsections.

\subsection{Vacuum equations}

According to our hypothesis,
in order to get the $U_4$ gravity equations we will assume that they
can be obtained from an Einstein-Hilbert  action using the scalar of
curvature (\ref{scurv}), the condition (\ref{pot}), and the
volume element (\ref{u4volume}),
\begin{eqnarray}
\label{vaction}
S_{\rm grav} &=& -\int d^4x e^{2\Theta} \sqrt{-g} \, R   \\
&=&-\int d^4x e^{2\Theta} \sqrt{-g} \left(
R^{V_4} + \frac{16}{3} \partial_\mu\Theta \partial^\mu \Theta
- \tilde{K}_{\nu\rho\alpha} \tilde{K}^{\alpha\nu\rho}
\right) + {\rm surface \ terms}. \nonumber
\end{eqnarray}
where the generalized Gauss' formula (\ref{gauss}) was used.

The equations for the $g^{\mu\nu}$, $\Theta$, and
$\tilde{K}_{\nu\rho\alpha}$ fields follow from the minimization of the action
(\ref{vaction}).
The variations of $g^{\mu\nu}$ and $S_{\mu\nu}^{\ \ \rho}$ are assumed to
vanish in the boundary.
The equation $\frac{\delta S_{\rm grav}}{\delta\tilde{K}_{\nu\rho\alpha}} =0$
implies that $\tilde{K}^{\nu\rho\alpha} =  0$,
$\frac{\delta S_{\rm grav}}{\delta\tilde{K}_{\nu\rho\alpha}}$ standing for the
Euler-Lagrange equations for
${\delta\tilde{K}_{\nu\rho\alpha}}$.
 For the other equations we have
\begin{eqnarray}
\label{1st}
-\frac{e^{-2\Theta}}{\sqrt{-g}}
\left.\frac{\delta }{\delta g^{\mu\nu}}S_{\rm grav}
\right|_{\tilde{K}=0} &=& R^{V_4}_{\mu\nu}
-2D_\mu \partial_\nu\Theta \nonumber \\
&&-\frac{1}{2}g_{\mu\nu}
\left(
R^{V_4} + \frac{8}{3}\partial_\rho\Theta \partial^\rho \Theta
-4 \Box \Theta
\right) = 0,   \\
-\frac{e^{-2\Theta}}{2\sqrt{-g}}
\left.\frac{\delta }{\delta \Theta}S_{\rm grav}
\right|_{\tilde{K}=0} &=&
R^{V_4} + \frac{16}{3}\left(
\partial_\mu\Theta \partial^\mu \Theta -
\Box \Theta \right) =0, \nonumber
\end{eqnarray}
where $R^{V_4}_{\mu\nu}$ is the $V_4$ Ricci tensor, calculated using the
Christoffel symbols, and $\Box$ is the $U_4$ d'Alambertian operator,
$\Box = D_\mu D^\mu$.

Taking the trace of the first equation of (\ref{1st}),
\begin{equation}
R^{V_4} + \frac{16}{3}\partial_\mu\Theta \partial^\mu \Theta =
6\Box\Theta,
\end{equation}
and using it,  one finally obtains
the new $U_4$
gravity equations for the vacuum,
\begin{eqnarray}
\label{vacum0}
R^{V_4}_{\mu\nu} &=& 2D_\mu\partial_\nu \Theta
- \frac{4}{3} g_{\mu\nu}\partial_\rho\Theta \partial^\rho \Theta
= 2D_\mu S_\nu - \frac{4}{3}g_{\mu\nu}S_\rho S^\rho, \nonumber \\
\Box \Theta &=& \frac{e^{-2\Theta}}{\sqrt{-g}}
\partial_\mu e^{2\Theta}\sqrt{-g}\partial^\mu\Theta = D_\mu S^\mu = 0, \\
\tilde{K}_{\alpha\beta\gamma} &=& 0. \nonumber
\end{eqnarray}

The vacuum equations (\ref{vacum0})
point out new features of EC theory. It is
clear that torsion, described by the last two equations,
 propagates.
The torsion mediated interactions are not of
contact type anymore. The traceless tensor $\tilde{K}_{\alpha\beta\gamma}$
is zero for the vacuum, and only the trace $S_\mu$ can be non-vanishing
outside matter distributions. As it is expected, the gravity field
configuration for the vacuum is determined only
by boundary conditions, and if
due to such conditions we have that $S_\mu=0$, our equations reduce to the
usual vacuum equations, $S_{\alpha\gamma\beta}=0$, and
$R^{V_4}_{\alpha\beta}=0$.

In a first sight, it seems that the equations
(\ref{vacum0}) can be written without using the scalar
potential $\Theta(x)$, so it can be questioned if they are valid for
manifolds not obeying (\ref{pot}). The answer is clearly negative, since
the first term in the right-handed side of the first equation
is symmetrical under the change $(\mu\leftrightarrow\nu)$ only if the condition
(\ref{pot}) holds.

Another remarkable consequence of the proposed model, is that the
action (\ref{vaction}) can provide a physical interpretation for the dilaton
gravity in terms of the non-riemannian structure of space time\cite{saa3}.
Remember that dilaton gravity arises as consequence of conformal invariance
in a quantum analysis up to one-loop order for string theory in background
fields\cite{callan,green}, and its equations can be gotten from the
following effective action:
\begin{equation}
\label{actiondil}
S = -\int d^Nx\sqrt{-g}e^{-2\Phi}\left(
R^{V_4} + 4\partial_\mu\Phi \partial^\mu\Phi -
\frac{1}{12}H_{\alpha\beta\gamma}
H^{\alpha\beta\gamma} - \frac{N-26}{3\alpha^\prime}
\right),
\end{equation}
where $N$ is the space-time dimension, $\Phi$ is the dilaton field, and
$H_{\alpha\beta\gamma}$ is the totally anti-symmetrical tensor, which is
 given by
\begin{equation}
\label{aga}
H_{\alpha\beta\gamma} = \partial_\alpha B_{\beta\gamma} +
\partial_\gamma B_{\alpha\beta} +\partial_\beta B_{\gamma\alpha},
\end{equation}
where $B_{\mu\nu}$ is the massless anti-symmetrical tensor background field.
It is well known from sigma model context, that $B_{\mu\nu}$ is related to the
torsion of space-time\cite{zachos}, but up to now there was no geometrical
interpretation for the dilaton field $\Phi$.

The similarity between (\ref{vaction}) and (\ref{actiondil}) is clear.
They can be identified if one assumes that:
\begin{eqnarray}
\label{ident}
\Theta(x) &=& -\Phi(x), \nonumber \\
\frac{1}{12}H_{\alpha\beta\gamma}H^{\alpha\beta\gamma} &=&
\tilde{K}_{\nu\rho\alpha} \tilde{K}^{\alpha\nu\rho}
- \frac{4}{3} \partial_\mu\Phi \partial^\mu \Phi =
{K}_{\nu\rho\alpha} {K}^{\alpha\nu\rho} \\
\Lambda &=&  -\frac{N-26}{3\alpha^\prime},,\nonumber
\end{eqnarray}
where $\Lambda$ stands for a cosmological constant, not present in the
original action (\ref{vaction}). The  second equation of (\ref{ident})
can be used to write the contorsion tensor in terms of the tensor
$H_{\alpha\beta\gamma}$,
which is consistent with the result that $H_{\alpha\beta\gamma}$ is related
to torsion\cite{zachos}. The case $\Phi=0$ is namely the prediction from
sigma models. The interpretation of $K_{\alpha\gamma\beta}$ in the usual
frame of EC gravity is problematical, since
the totally anti-symmetrical
tensor $H_{\alpha\beta\gamma}$ is invariant under the ``gauge'' transformation
\begin{equation}
B_{\mu\nu}\rightarrow B_{\mu\nu} +  \partial_\mu\Lambda_\nu -
\partial_\nu\Lambda_\mu,
\end{equation}
that is a consequence of this definition (\ref{aga}), and
there are no reasons {\em a priori}
to expect such behavior of the contorsion tensor, or what is equivalent,
it is not expected in the frame of EC theory
that the contorsion tensor can be derived from an
anti-symmetrical field like  $B_{\mu\nu}$ in (\ref{aga}). However, it is
clear  the relation between the dilaton field $\Phi$ and
the potential for the trace of the torsion tensor $\Theta$, which allow us
to interpret the dilaton field as being part of the non-riemannian
structure of space-time.

\subsection{Scalar fields}

The first step to introduce matter fields in our discussion
will be the description of
scalar fields on RC manifolds.
In order to do it, we will use minimal coupling procedure
(MCP), which consists; given a Lorentz invariant action, to change the
usual derivatives by covariant ones, the Lorentz metric tensor by the general
one, and to introduce the  covariant volume element, that according
to our hypothesis will be given by (\ref{u4volume}).
Using MCP for a massless scalar field one gets
\begin{eqnarray}
\label{scala}
S = S_{\rm grav} + S_{\rm scal} &=& -\int  \,d^4xe^{2\Theta}\sqrt{-g}
\left(R -\frac{g^{\mu\nu}}{2} \partial_\mu\varphi \partial_\nu \varphi
\right)\\
&=&-\int d^4x e^{2\Theta} \sqrt{-g} \left(
R^{V_4} + \frac{16}{3} \partial_\mu\Theta \partial^\mu \Theta
- \tilde{K}_{\nu\rho\alpha} \tilde{K}^{\alpha\nu\rho}
-\frac{g^{\mu\nu}}{2} \partial_\mu\varphi \partial_\nu \varphi
\right), \nonumber
\end{eqnarray}
where surface terms were discarded.
The equations for this case are obtained by varying (\ref{scala}) with
respect to $\varphi$, $g^{\mu\nu}$, $\Theta$, and
$\tilde{K}_{\alpha\beta\gamma}$. As in the vacuum case, the equation
$\frac{\delta S}{\delta \tilde{K}}=0$
implies $\tilde{K}=0$. Taking it into
account we have
\begin{eqnarray}
\label{e1}
-\frac{e^{-2\Theta}}{\sqrt{-g}} \left.
\frac{\delta S}{\delta\varphi}
\right|_{\tilde{K}=0} &=&  \frac{e^{-2\Theta}}{\sqrt{-g}}\partial_\mu
e^{2\Theta}\sqrt{-g}\partial^\mu\varphi
=\Box \varphi = 0, \nonumber \\
-\frac{e^{-2\Theta}}{\sqrt{-g}} \left.
\frac{\delta S}{\delta g^{\mu\nu}}
\right|_{\tilde{K}=0} &=& R^{V_4}_{\mu\nu}
- 2 D_\mu S_\nu - \frac{1}{2} g_{\mu\nu}
\left(
R^{V_4} + \frac{8}{3}S_\rho S^\rho - 4 D_\rho S^\rho
\right) \nonumber \\
&&-\frac{1}{2} \partial_\mu \varphi \partial_\nu\varphi
+ \frac{1}{4} g_{\mu\nu}\partial_\rho \varphi \partial^\rho \varphi = 0, \\
-\frac{e^{-2\Theta}}{2\sqrt{-g}} \left.
\frac{\delta S}{\delta \Theta}
\right|_{\tilde{K}=0} &=& R^{V_4} +
\frac{16}{3}\left( S_\mu S^\mu - D_\mu S^\mu\right)
 -\frac{1}{2} \partial_\mu\varphi \partial^\mu\varphi = 0. \nonumber
\end{eqnarray}
Taking the trace of the second equation of (\ref{e1}),
\begin{equation}
R^{V_4} + \frac{16}{3} S_\mu S^\mu = 6 D_\mu S^\mu +
\frac{1}{2} \partial_\mu\varphi \partial^\mu \varphi,
\end{equation}
and using it, we get the following
set of equations for the massless scalar case
\begin{eqnarray}
\label{aa}
\Box \varphi &=&  0, \nonumber \\
R^{V_4}_{\mu\nu} &=& 2D_\mu S_\nu - \frac{4}{3}g_{\mu\nu} S_\rho S^\rho
+\frac{1}{2} \partial_\mu\varphi \partial_\nu\varphi, \\
D_\mu S^\mu &=& 0, \nonumber \\
\tilde{K}_{\alpha\beta\gamma} &=& 0. \nonumber
\end{eqnarray}

As one can see, the torsion equations have the same form than the ones
 of the vacuum case, (\ref{vacum0}). Any
contribution to the torsion will be due to boundary conditions, and not due
to the scalar field itself.
It means that if such boundary conditions imply that $S_\mu=0$, the
equations for the fields $\varphi$ and $g_{\mu\nu}$ will be the same ones
of the general relativity.
One can interpret this by saying that,
even feeling the torsion (see the second equation of (\ref{aa})),
massless scalar fields do not produce it. Such  behavior is
compatible with the idea that torsion must be governed by spin distributions.

However, considering massive scalar fields,
\begin{eqnarray}
S_{\rm scal} = \int  \,d^4xe^{2\Theta}\sqrt{-g}
\left(\frac{g^{\mu\nu}}{2} \partial_\mu\varphi \partial_\nu \varphi
-\frac{m^2}{2}\varphi^2 \right),
\end{eqnarray}
 we have the
following set of equations instead of (\ref{aa})
\begin{eqnarray}
\label{aa1}
(\Box+m^2) \varphi &=&  0, \nonumber \\
R^{V_4}_{\mu\nu} &=& 2D_\mu S_\nu - \frac{4}{3}g_{\mu\nu} S_\rho S^\rho
+\frac{1}{2} \partial_\mu\varphi \partial_\nu\varphi
-\frac{1}{2} m^2\varphi^2, \\
D_\mu S^\mu &=& \frac{3}{4}m^2\varphi^2, \nonumber \\
\tilde{K}_{\alpha\beta\gamma} &=& 0. \nonumber
\end{eqnarray}
The equation for the trace of the torsion tensor is different than the one of
the vacuum case, we have that massive scalar fields can produce torsion.
In contrast to the massless case, the equations (\ref{aa1}) do not admit as
solution $S_\mu=0$ for non-vanishing $\varphi$.
This is in disagreement with the traditional belief that torsion must be
governed by spin distributions. We will return to this point in the last
section.

\subsection{Gauge fields}

We need to be careful with the use of MCP to gauge fields. We will restrict
ourselves to the abelian case in this work,
 non-abelian gauge fields will bring some
technical difficulties that  will not contribute to the understanding
of the basic problems  of gauge fields on Riemann-Cartan space-times.

It is well known that Maxwell field can be described by the differential
$2$-form
\begin{equation}
F = dA = d(A_\alpha dx^\alpha) = \frac{1}{2}F_{\alpha\beta}dx^\alpha
\label{form}
\wedge dx^\beta,
\end{equation}
where $A$ is the (local) potential $1$-form, and
$F_{\alpha\beta}=\partial_\alpha A_\beta- \partial_\beta A_\alpha$ is the
usual electromagnetic tensor. It is important to stress that the
forms $F$ and
$A$ are covariant objects in any differentiable manifolds. Maxwell's equations
can be written in Minkowski space-time in terms of exterior calculus as
\begin{eqnarray}
\label{maxeq}
dF&=&0, \\
d {}^*\!F &=& 4\pi {}^*\! J, \nonumber
\end{eqnarray}
where ${}^*$ stands for the Hodge star operator and $J$ is the current
$1$-form, $J=J_\alpha dx^\alpha$. The first equation in (\ref{maxeq}) is
a consequence of the definition (\ref{form}) and of Poincar\'e's lemma.
In terms of components, one has the familiar homogeneous and non-homogeneous
Maxwell's equations,
\begin{eqnarray}
\label{maxeq1}
\partial_{[\gamma} F_{\alpha\beta]} &=& 0, \\
\partial_\mu F^{\nu\mu} &=& 4\pi J^\nu, \nonumber
\end{eqnarray}
where ${}_{[\ \ \ ]}$ means antisymmetrization. We know also that the
non-homogenous equation follows from the minimization
of the following action
\begin{equation}
S = -\int \left(4\pi{}^*\!J\wedge A +\frac{1}{2} F \wedge {}^*\!F\right) =
    \int d^4x\left(4\pi J^\alpha A_\alpha - \frac{1}{4}
F_{\alpha\beta}F^{\alpha\beta} \right).
\label{actmink}
\end{equation}

If one tries to cast (\ref{actmink}) in a covariant way by using MCP in the
tensorial quantities, we have that Maxwell tensor will be given by
\begin{equation}
\label{tilda}
F_{\alpha\beta}\rightarrow
\tilde{F}_{\alpha\beta} =
F_{\alpha\beta} - 2 S_{\alpha\beta}^{\ \ \rho}A_\rho,
\end{equation}
which explicitly breaks gauge invariance. With this analysis, it usually
arises the conclusion that gauge fields cannot interact minimally with
Einstein-Cartan gravity. We would like also to stress another undesired
consequence, also related to the losing of gauge symmetry, of the use of MCP
in the tensorial quantities. The homogeneous Maxwell's equation, the
first of (\ref{maxeq1}), does not come from a Lagrangian, and of course,
if we choose to use
MCP in the tensorial quantities we need also apply MCP to it. We get
\begin{equation}
\partial_{[\alpha} \tilde{F}_{\beta\gamma]} +
2 S_{[\alpha\beta}^{\ \ \rho} \tilde{F}_{\gamma]\rho} = 0 ,
\label{falac}
\end{equation}
where $\tilde{F}_{\alpha\beta}$ is given by (\ref{tilda}). One can see that
(\ref{falac}) has no general solution for arbitrary
$S_{\alpha\beta}^{\ \ \rho}$. Besides the losing of gauge symmetry,
the use o MCP in the tensorial quantities also leads to a non consistent
homogeneous equation. One has enough arguments to avoid the use of
MCP in the tensorial quantities for gauge fields.

However, MCP can be successfully applied for general gauge fields
(abelian or not) in the differential form quantities \cite{saa2}.  As its
consequence, one has that the homogeneous equations is already in a
covariant form in any differentiable manifold, and that the covariant
non-homogeneous equations can be gotten from a Lagrangian obtained only by
changing the metric tensor and by
introducing the ``parallel'' volume element in the Minkowskian action
(\ref{actmink}). Considering the case where $J^\mu=0$, we have the
following action to describe the interaction of Maxwell fields and
Einstein-Cartan gravity
\begin{equation}
\label{actmax}
S = S_{\rm grav} + S_{\rm Maxw} = -\int   \,d^4x e^{2\Theta} \sqrt{-g}
\left(
R + \frac{1}{4}F_{\mu\nu}F^{\mu\nu}
\right).
\end{equation}
As in the previous cases, the equation $\tilde{K}_{\alpha\beta\gamma}=0$
follows from the minimization of (\ref{actmax}).
The other equations will be
\begin{eqnarray}
\label{ee1}
&&\frac{e^{-2\Theta}}{\sqrt{-g}}\partial_\mu e^{2\Theta}\sqrt{-g} F^{\nu\mu}
=0, \nonumber \\
&& R_{\mu\nu}^{V_4} = 2D_\mu S_\nu - \frac{4}{3}g_{\mu\nu}S_\rho S^\rho
-\frac{1}{2} \left(F_{\mu\alpha}F^{\ \alpha}_\nu
+\frac{1}{2}g_{\mu\nu} F_{\omega\rho}F^{\omega\rho} \right), \\
&& D_\mu S^\mu = -\frac{3}{8}F_{\mu\nu}F^{\mu\nu}. \nonumber
\end{eqnarray}

One can see that the equations (\ref{ee1}) are invariant under the usual
$U(1)$ gauge transformations. It is also clear
from the equations (\ref{ee1}) that Maxwell fields interact with the
non-Riemannian structure of space-time. We have that, as in the massive
scalar case, the equations do not admit as solution $S_\mu=0$ for arbitrary
$F_{\alpha\beta}$, Maxwell fields are also sources to the space-time torsion.
Similar results can be obtained also for non-abelian gauge fields\cite{saa2}.

\subsection{Fermion fields}

The Lagrangian for a (Dirac)
fermion field with mass $m$ in the Minkowski space-time
is given by
\begin{equation}
{\cal L}_{\rm F}=\frac{i}{2}\left(\overline{\psi}\gamma^a\partial_a\psi
- \left(\partial_a\overline{\psi} \right)\gamma^a\psi \right)
- m\overline{\psi}\psi,
\label{fermion}
\end{equation}
where $\gamma^a$ are the Dirac matrices and
$\overline{\psi}=\psi^\dagger\gamma^0$. Greek indices denote space-time
coordinates (holonomic), and roman ones locally flat coordinates
(non-holonomic). It is well known\cite{hehl,venzo}
that in order to cast (\ref{fermion}) in a covariant way, one needs to
introduce the vierbein field, $e^\mu_a(x)$, and
to generalize the Dirac matrices,\break
$\gamma^\mu(x) = e^\mu_a(x)\gamma^a$. The partial derivatives also must be
generalized with the introduction of the spinorial connection $\omega_\mu$,
\begin{eqnarray}
\partial_\mu\psi \rightarrow
 \nabla_\mu\psi &=& \partial_\mu\psi+ \omega_\mu \psi, \nonumber \\
\partial_\mu\overline{\psi} \rightarrow
 \nabla_\mu\overline{\psi} &=& \partial_\mu\overline{\psi} -
\overline{\psi}\omega_\mu,
\end{eqnarray}
where the spinorial connection is given by
\begin{eqnarray}
\label{spincon}
\omega_\mu &=& \frac{1}{8}[\gamma^a,\gamma^b]e^\nu_a\left(
\partial_\mu e_{\nu b} -\Gamma^\rho_{\mu\nu}e_{\rho b}\right) \\
&=& \frac{1}{8}\left(
\gamma^\nu\partial_\mu\gamma_\nu - \left(\partial_\mu\gamma_\nu \right)
\gamma^\nu - \left[\gamma^\nu,\gamma_\rho \right] \Gamma^\rho_{\mu\nu}
\right). \nonumber
\end{eqnarray}
 The last
step, according to our hypothesis, shall be
the introduction of the ``parallel''
 volume element, and after that one
gets the following action for fermion fields on RC manifolds
\begin{equation}
S_{\rm F} = \int d^4x e^{2\Theta}\sqrt{-g}\left\{
\frac{i}{2}\left(\overline{\psi}\gamma^\mu(x)\nabla_\mu\psi -
\left(\nabla_\mu\overline{\psi}\right)\gamma^\mu(x)\psi \right)
-m\overline{\psi}\psi
\right\}.
\label{fermioncov}
\end{equation}

Varying the action (\ref{fermioncov}) with respect to $\overline{\psi}$ one
obtains:
\begin{equation}
\frac{e^{-2\Theta}}{\sqrt{-g}}\frac{\delta S_{\rm F}}{\delta\overline{\psi}} =
\frac{i}{2}\left(\gamma^\mu\nabla_\mu\psi + \omega_\mu\gamma^\mu\psi \right)
-m \psi + \frac{i}{2}\frac{e^{-2\Theta}}{\sqrt{-g}} \partial_\mu
e^{2\Theta}\sqrt{-g}\gamma^\mu\psi = 0.
\end{equation}
Using the result
\begin{equation}
[\omega_\mu,\gamma^\mu]\psi = - \left(
\frac{e^{-2\Theta}}{\sqrt{-g}}\partial_\mu e^{2\Theta}\sqrt{-g}\gamma^\mu
\right)\psi,
\end{equation}
that can be check using  (\ref{spincon}),
(\ref{gammacontr}),  and
properties of ordinary Dirac matrices and of the vierbein field,
 we get the following equation for $\psi$ on a RC space-time:
\begin{equation}
\label{psi}
i\gamma^\mu(x)\nabla_\mu\psi - m\psi =0.
\end{equation}
The equation for $\overline{\psi}$ can be obtained in a similar way,
\begin{equation}
\label{psibar}
i \left( \nabla_\mu\overline{\psi}\right) \gamma^\mu(x)
 + m\overline{\psi} = 0.
\end{equation}

We can see that the equations (\ref{psi}) and (\ref{psibar}) are the same
ones that arise from MCP used in the minkowskian equations of motion. In the
usual EC theory, the equations obtained from the action principle do not
coincide with the equations gotten by generalizing the minkowskian
ones\cite{venzo}. This is another new feature of the proposed model, it is
more consistent in the sense that
it makes no difference if one starts by the
equations of motion or by the minimal action principle.

The Lagrangian that describes the interaction of fermion fields with the
Einstein-Cartan gravity is
\begin{eqnarray}
\label{actferm}
S &=& S_{\rm grav} +
S_{\rm F} \\ &=& - \int d^4x e^{2\Theta}\sqrt{-g} \left\{
R - \frac{i}{2}\left(\overline{\psi}\gamma^\mu\partial_\mu\psi -
\left(\partial_\mu\overline{\psi}\right)\gamma^\mu\psi +
\overline{\psi}\left[\gamma^\mu,\omega_\mu\right] \psi\right)
  + m\overline{\psi}\psi \right\} \nonumber \\
&=& - \int d^4x e^{2\Theta}\sqrt{-g} \left\{
R - \frac{i}{2}\left(\overline{\psi}\gamma^\mu\partial_\mu\psi -
\left(\partial_\mu\overline{\psi}\right)\gamma^\mu\psi +
\overline{\psi}\left[\gamma^\mu,\omega^{V_4}_\mu\right] \psi\right)
\right.\nonumber \\
&&\ \ \ \ \ \ \ \ \ \ \
\left. -\frac{i}{8}\overline{\psi}\tilde{K}_{\mu\nu\omega}
\gamma^{[\mu}\gamma^\nu\gamma^{\omega]} \psi
  + m\overline{\psi}\psi \right\},\nonumber
\end{eqnarray}
where it was used that
$\gamma^a\left[\gamma^b,\gamma^c\right]+
\left[\gamma^b,\gamma^c\right]\gamma^a=
2\gamma^{[a}\gamma^b\gamma^{c]}$, and that
\begin{equation}
\omega_\mu = \omega^{V_4}_\mu +\frac{1}{8}K_{\mu\nu\rho}
\left[\gamma^\nu,\gamma^\rho\right],
\end{equation}
where $\omega^{V_4}_\mu$ is the $V_4$ spinorial connection, calculated by
using the Christoffel symbols instead of the full connection in
(\ref{spincon}).

The peculiarity of fermion fields is that one has no trivial equation
for $\tilde{K}$ from (\ref{actferm}).
The Euler-Lagrange equations for $\tilde{K}$ is given by
\begin{eqnarray}
\frac{e^{-2\Theta}}{\sqrt{-g}} \frac{\delta S}{\delta\tilde{K}}  =
\tilde{K}^{\mu\nu\omega} + \frac{i}{8}\overline{\psi}
\gamma^{[\mu}\gamma^\nu\gamma^{\omega]}\psi = 0.
\label{ka}
\end{eqnarray}
Differently from the previous cases, we have that the  traceless part of
the contorsion tensor,
$\tilde{K}_{\alpha\beta\gamma}$, is proportional to the spin
distribution. It is still zero outside
matter distribution, since its equation is an algebraic one, it does not
allow propagation. The other equations follow from the minimization of
(\ref{actferm}). The main difference between these equations and the usual
ones obtained from standard EC gravity, is that in the present case one
has non-trivial solution for the trace of the torsion tensor, that is
derived from $\Theta$. In the standard EC gravity, the torsion tensor is
a totally anti-symmetrical tensor, its equation is identical to
(\ref{ka}), and so it has vanishing trace.

\section{Final remarks}

In this section, we are going to discuss the role of the
 condition (\ref{pot}), and to show that for the proposed model, the
traditional point of view for the spin--torsion relationship must be
modified. First of all, we would like to stress that, in contrast to the
standard EC theory of gravity, due to the introduction of the new volume
element, it makes no difference to use properly MCP in the equations of
motion or in the action formulation. This is a strong difference between
the standard EC theory and general relativity. We know that in general
relativity, using MCP in the equations of motion or in the action, we
get the same result.

It was already said the condition that the trace of the torsion
tensor is derive from a scalar potential, (\ref{pot}), is the necessary
condition in order to be possible the definition of a connection-compatible
volume element on a manifold. Therefore, we have that our approach is
restrict to space-times which admits such volume elements. However, assuming
that one needs to use connection-compatible volume elements in action
formulations, we automatic have this restriction if we wish to use
minimal action principles.

In spite of it is not clear how to get EC gravity equations without
using a minimal action principle, we can speculate about matter fields
on space-times not obeying (\ref{pot}). Because of it is equivalent
to use MCP in the equations of motion or in the action formulation, we
can forget the last and to cast the equations of motion for matter
fields in a covariant way directly. It can be done easily, as example,
for scalar fields\cite{saa1}. We get the following equation
\begin{eqnarray}
\label{spscal}
\partial_\mu\partial^\mu\varphi =0 \rightarrow
\frac{1}{\sqrt{-g}}\partial_\mu\sqrt{-g}\partial^\mu\varphi
 + 2S_\mu\partial^\mu\varphi=0,
\end{eqnarray}
which is, apparently, a consistent equation. However, we need to define
a inner product for the space of the solutions of (\ref{spscal})
\cite{dewitt}, and we are able to do it only if (\ref{pot}) holds.
We have that the dynamics of matter fields requires some restrictions
to the non-riemannian structure of space-time, namely, the condition
(\ref{pot}). This is more evident for gauge fields, where
(\ref{pot}) arises directly as an integrability condition for the
equations of motion \cite{saa2}. It seems that condition (\ref{pot}) cannot
be avoided.

We could realize from the matter fields studied, that the trace of the
torsion tensor is not directly related to spin distributions. This is a
new feature of the proposed model, and we naturally arrive to the
following question: What is the source of torsion? The situation for the
traceless part of the torsion tensor is the same that one has in the
standard EC theory, only fermion fields can be sources to it. For the
trace, it is quite different.
 Take for example $\tilde{K}_{\alpha\beta\gamma}=0$, that corresponds to
scalar and gauge fields.
In these cases, the equation
for the trace of the torsion tensor is given by
\begin{equation}
\label{ds}
D_\mu S^\mu = \frac{3}{2}\frac{e^{-2\Theta}}{\sqrt{-g}}
\left(
g^{\mu\nu}\frac{\delta S_{\rm mat}}{\delta g^{\mu\nu}} + \frac{1}{2}
\frac{\delta S_{\rm mat}}{\delta \Theta}
\right).
\end{equation}
Using the definition of the energy-momentum tensor
\begin{equation}
\frac{e^{-2\Theta}}{\sqrt{-g}}
\frac{\delta S_{\rm mat}}{\delta g^{\mu\nu}} = -\frac{1}{2}T_{\mu\nu},
\end{equation}
and that for scalar and gauge fields we have
\begin{equation}
\frac{e^{-2\Theta}}{\sqrt{-g}}
\frac{\delta S_{\rm mat}}{\delta \Theta} = 2 {\cal L}_{\rm mat},
\end{equation}
equation (\ref{ds}) leads to
\begin{equation}
D_\mu S^\mu = \frac{3}{2}
\left(  {\cal L}_{\rm mat} - \frac{1}{2}T
\right),
\end{equation}
where $T$ is the trace of the energy-momentum tensor.
The quantity between parenthesis, in general, has nothing to do with spin, and
in spite of this, it is the source for a part of the torsion.
For gauge fields, for which $T=0$, the source for the trace of the torsion
tensor is namely the Lagrangian.
The presence
of the scalar potential $\Theta$ in the ``parallel'' $U_4$ volume
element seems to indicate that torsion is not directly related to spin
distributions. These topics are now under investigation.

\acknowledgements

The author is grateful to Professor Josif Frenkel. This work was supported
by Funda\c c\~ao de Amparo \`a Pesquisa do Estado de S\~ao Paulo.

\end{document}